\documentclass[a4paper,twocolumn,english,american,prb,fixfloat,notitlepage]{revtex4-1}
\usepackage[latin9]{inputenc}
\usepackage{xcolor}
\usepackage{amsmath}
\usepackage{amssymb}
\usepackage{graphicx}
\PassOptionsToPackage{normalem}{ulem}
\usepackage{ulem}

\makeatletter

\pdfpageheight\paperheight
\pdfpagewidth\paperwidth

\providecolor{lyxadded}{rgb}{0,0,1}
\providecolor{lyxdeleted}{rgb}{1,0,0}
\DeclareRobustCommand{\mklyxadded}[1]{\bgroup\color{lyxadded}{}#1\egroup}
\DeclareRobustCommand{\mklyxdeleted}[1]{\bgroup\color{lyxdeleted}\mklyxsout{#1}\egroup}
\DeclareRobustCommand{\mklyxsout}[1]{\ifx\\#1\else\sout{#1}\fi}
\usepackage{xcolor}
\definecolor{darkblue}{rgb}{0.1,0.2,0.6} 
\definecolor{lightblue}{rgb}{0.1,0.1,1.0}
\definecolor{darkred}{rgb}{0.8,0.1,0.2}
\usepackage[colorlinks,citecolor=lightblue,linkcolor=darkblue,urlcolor=lightblue] {hyperref}

\makeatother

\usepackage{babel}
\begin{document}
\title{Dynamic scaling relation in quantum many-body systems}
\author{Devendra Singh Bhakuni}
\affiliation{Department of Physics, Ben-Gurion University of the Negev, Beer-Sheva
84105, Israel}
\author{Yevgeny Bar~Lev}
\affiliation{Department of Physics, Ben-Gurion University of the Negev, Beer-Sheva
84105, Israel}
\email{ybarlev@bgu.ac.il}

\begin{abstract}
In delocalized systems, particle number fluctuations, also known as
quantum surface roughness, and the mean-square displacement exhibit
a temporal power-law growth followed by a saturation to a system-size-dependent
value. We use simple scaling arguments to show that these quantities
satisfy the Family-Vicsek scaling law and derive a dynamic scaling
relation between the dynamical exponents, assuming that the saturation
times of both quantities scale similarly with the system size. This
relation clarifies the mechanism behind quantum surface roughness
growth and suggests that diffusive quantum many-body systems belong
to the Edwards-Wilkinson universality class. Moreover, it provides
a convenient way to assess quantum transport in cold-atoms experiments.
We numerically verify our results by studying two non-interacting
models and one interacting model having regimes with distinct dynamical
exponents.
\end{abstract}
\maketitle
\section{Introduction}
In the vicinity of a continuous phase transition,
many physical properties of a system exhibit a power-law dependence,
manifested by critical exponents~\citep{stanley1971phase,zinnjustin2002quantum,odor_universality_2004}.
Scaling arguments show that the critical exponents are not independent,
but related via scaling relations~\citep{zinnjustin2002quantum,odor_universality_2004}.
Moreover, renormalization group theory explains why microscopically
different physical systems, which belong to the same universality
class, share the same critical exponents~\citep{stanley1971phase,odor_universality_2004,hung_observation_2011,lesne_scale_2012,andreev_scale_2014}.
While the concepts of scaling and universality were originally introduced
for systems at equilibrium, they were successfully generalized to
classical out-of-equilibrium systems~\citep{broadbent1957percolation,edwards_surface_1982,vicsek_dynamic_1984,family_scaling_1985,kinzel1985,kardar_dynamic_1986,barabasi1995fractal,bray1994theory,halpinhealy1995215,krug1997origins,Haye2000nonequilibrium,takeuchi2011,physrevlett.104.230601,corwin_kardar-parisi-zhang_2016}.
One of the prominent examples of dynamical scaling occurs in classical
surface physics~\citep{edwards_surface_1982,vicsek_dynamic_1984,family_scaling_1985}.
Surface roughness of a surface segment of length $L$ is defined as
the standard deviation of surface height. It typically increases as
a power law in time before saturating to a value which depends on
the surface segment length, following the famous Family-Vicsek (FV)
scaling~\citep{vicsek_dynamic_1984,family_scaling_1985}, 
\begin{align}
w\left(L,t\right)= & L^{\alpha}f\left(t/L^{z}\right),\label{eq:fv_scaling}
\end{align}
where $f\left(x\right)$ is a unitless function which for $t\ll L^{z}$
grows as $f\left(x\right)\sim x^{\beta}$ and for $t\gg L^{z}$ saturates
to a constant independent of $L$. The exponents $\alpha$ and $\beta$
are the roughening and growth exponents, respectively, and the exponent
$z,$ called a dynamical exponent defines the saturation time, $t_{\text{sat}}\sim L^{z}$.
Since the early time growth of $w\left(L,t\right)$ does not depend
on the system size, from the properties of $f\left(x\right)$, it
follows that the exponents satisfy the dynamic scaling relation $z=\alpha/\beta$.
Well-known universality classes in one-dimensional classical systems
are the Kardar-Parisi-Zhang (KPZ)~\citep{kardar_dynamic_1986} class
with $\alpha=1/2$ and $\beta=1/3$ and the Edwards-Wilkinson (EW)
class with $\alpha=1/2$ and $\beta=1/4$~\citep{edwards_surface_1982}.
In higher dimensions, the solution of the EW equation suggests $\alpha,\beta\leq0$~\citep{edwards_surface_1982,Nattermann1992kinetic},
while numerical studies have shown logarithmic growth~\citep{pal_dynamical_2003,pal_edwardswilkinson_1999,plischke_time-reversal_1987,kwak_random_2019}.
The KPZ class in higher dimensions was also studied numerically~\citep{oliveira_kardar-parisi-zhang_2013,pagnani_numerical_2015}.
These classes have found application in a wide variety of physical
systems such as sedimentation of colloidal particles~\citep{mccloud_deposition_1997},
growth of bacterial colonies~\citep{vicsek_self-affine_1990}, fire
fronts~\citep{maunuksela_kinetic_1997,myllys_kinetic_2001}, spin
chains~\citep{de_nardis_universality_2020,dupont_universal_2020,jin_stochastic_2020,cai_1_2022,cecile2023squeezed},
and driven-dissipative condensates~\citep{Squizzato_kardar_2018,Deligiannis_kardar_2022,fontaine_kardarparisizhang_2022}.

\begin{figure}
\includegraphics[width=0.7\columnwidth]{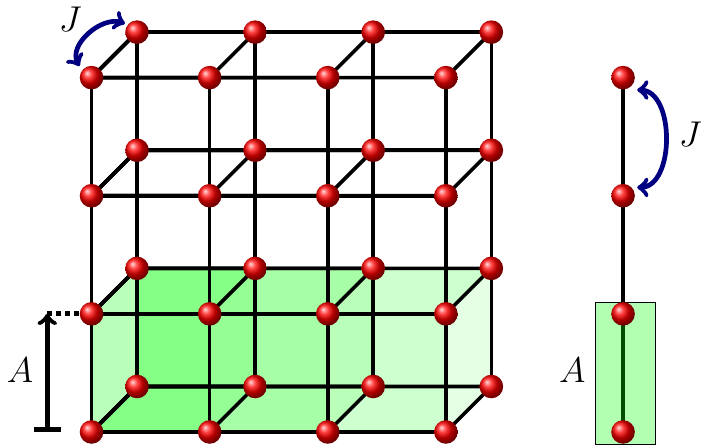}

\caption{Schematic of the domain in three-dimensions (left) and in one dimension
(right) used to calculate the surface roughness.}

\label{fig1}
\end{figure}

Quite recently, the notion of surface roughness was generalized to
quantum systems~\citep{fujimoto_dynamical_2021,fujimoto_family-vicsek_2020,fujimoto_impact_2022}.
Using an analogy between the fluctuating hydrodynamics and stochastic
surface growth, the \emph{quantum surface height} was defined as the
number of particles in a finite domain, such that, the quantum surface
roughness corresponds to the particle number fluctuations~\citep{spohn2014,das2014numerical,spohn2015,spohn2016}.
For delocalized non-interacting fermions the surface roughness was
shown to follow FV scaling with the exponents $\alpha=1/2$ and $\beta=1/2$,
which was called a ballistic class~\citep{fujimoto_dynamical_2021}.
In contrast, in the localized phase the FV scaling is not satisfied
due to suppression of roughness growth. It was argued that generic
delocalized systems will feature $\alpha=1/2$, while for systems
with a mixture of localized and delocalized states were shown to have
$\alpha,\ \beta\neq1/2$ \citep{fujimoto_dynamical_2021}. The question
what determines the growth exponent $\beta$ is however, still largely
open.

Another prominent dynamical quantity which shows power-law growth
is the spreading of density excitation of conserved quantities. In
delocalized systems, the spreading is typically governed by a power-law
with dynamical exponent, $\beta_{\text{tr}}$, followed by a saturation
to a system size dependent value. For single-particle systems, this
dynamical exponent is known to be related to the fractal dimension
of the single-particle eigenstates~\citep{ketzmerick_slow_1992,ketzmerick_what_1997,tomi_ohtsuki1997,luitz_multifractality_2020}.
In this paper, we establish a dynamic scaling relation between the
growth exponent of the surface roughness, or analogously, the particle
number fluctuations, to the dynamical exponent characterizing transport
of a corresponding conserve quantity. We demonstrate the relationship
by numerically studying two non-interacting models with tunable transport
regimes: the one-dimensional Fibonacci chain and the three-dimensional
Anderson model. Furthermore, we show that such a relationship also
holds for an interacting system.

\section{Surface roughness and particle fluctuations}\textit{\emph{The
total number of particles (fermions or bosons) in a finite connected
domain $A$ of a $d-$dimensional lattice is,}}
\begin{align}
\hat{N}_{A}= & \sum_{\boldsymbol{i}\in A}\hat{n}_{\boldsymbol{i}}\left(t\right),\label{eq:height_1d}
\end{align}
where $\hat{n}_{\boldsymbol{j}}$ is the particle number operator
on site \textbf{$\boldsymbol{j}$}. We note that up to an insignificant
constant this corresponds to the\emph{ quantum} surface height operator
as defined in Refs.~\citep{fujimoto_family-vicsek_2020,fujimoto_dynamical_2021}
(see appendix~\ref{particle-number-roughness}). We will designate the quantum
expectation of the number of particles as $\left\langle \hat{N}_{A}\left(t\right)\right\rangle =\text{Tr}\left[\hat{\rho}\left(t\right)\hat{N}_{A}\right]$,
where $\hat{\rho}\left(t\right)$ is the density matrix of the system
at time $t$.

The fluctuations of the particle number in the domain $A$ are given
by,
\begin{align}
\Delta N_{A}\left(t\right)= & \sqrt{\left\langle \hat{N}_{A}^{2}\left(t\right)\right\rangle -\left\langle \hat{N}_{A}\left(t\right)\right\rangle ^{2}}.\label{eq:particle-fluctuation}
\end{align}
Without the loss of generality, we will focus on a rectangular domain
$A$ which corresponds to a half of the system (see Fig.~\ref{fig1}).
We assume that the total number of particles is conserved, and the
continuity equation insures that the change in $N_{A}$ and $\Delta N_{A}$
is proportional to the surface area, $\left|\partial A\right|=L^{d-1}$,
of the domain $A$. In analogy to the classical surface roughness,
whose early time behavior does not depend on the dimensions of the
domain $A$, we normalize the particle fluctuations,
\begin{equation}
w\left(L,t\right)=\Delta N_{A}\left(t\right)/L^{\left(d-1\right)/2},\label{eq:surface_roughness}
\end{equation}
such that it corresponds to the definition of the quantum surface
roughness, extended to an arbitrary dimension (see Refs.~\citep{fujimoto_family-vicsek_2020,fujimoto_dynamical_2021},
and appendix \ref{particle-number-roughness}). Here, $L$ is the linear
dimension of the system, and for our choice of $A$, also the linear
dimension of $A$.

For one-dimensional delocalized systems, $w\left(L,t\right)$ follows
the FV scaling (Eq. \ref{eq:fv_scaling}) with roughening exponent $\alpha=1/2$
and a growth exponent $\beta$ which is system dependent \citep{fujimoto_family-vicsek_2020,fujimoto_dynamical_2021}
. In what follows we show, that this is also true in higher dimensions.
For thermalizing systems, initial states with sufficiently high energy
density, have exponentially decaying correlations between the positions
of the particles. Neglecting those correlations all together, which
is exact in the infinite temperature state, yields $w_{\infty}\left(L,t\right)\sim L^{1/2}$,
for any particle density and dimension (see appendix~\ref{particle-number-roughness}).
Comparing with (\ref{eq:fv_scaling}), this corresponds to $\alpha=1/2$.
For \emph{classical,} non-interacting diffusive systems the particle
fluctuations (\ref{eq:particle-fluctuation}) grow as $t^{1/4}$,
such that $\beta=1/4$ \citep{Kipnis1982,krapivsky2012fluctuations,eldad2022inverse}.

\section{Spreading of excitations and dynamical scaling relation} Transport of conserved quantities
at temperature $T$ can be characterized by the density-density correlation
function, 
\begin{equation}
C_{\boldsymbol{i}}\left(t\right)=\left\langle \left(\hat{n}_{\boldsymbol{i}}\left(t\right)-\left\langle \hat{n}_{\boldsymbol{i}}\right\rangle \right)\left(\hat{n}_{\boldsymbol{i_{0}}}\left(0\right)-\left\langle \hat{n}_{\boldsymbol{i}_{0}}\right\rangle \right)\right\rangle ,\label{eq:corr_function}
\end{equation}
where the expectation $\left\langle \hat{O}\right\rangle $ is taken
with respect to the finite temperature density matrix, $\hat{\rho}_{T}$,
such that $\left\langle \hat{n}_{\boldsymbol{i}}\left(t\right)\right\rangle \equiv\text{Tr\ensuremath{\hat{\rho}_{T}\hat{n}_{\boldsymbol{i}}\left(t\right)}}$.
The correlation function describes the spreading of a density excitation
at site $\boldsymbol{i}_{0}$ of a $d$-dimensional lattice~\citep{steinigeweg_2009,bar_lev_absence_2015,steinigeweg_real-time_2017,luitz_ergodic_2017}.
The width of this excitation, also called the mean-squared displacement
(MSD), is given by
\begin{equation}
R^{2}\left(t\right)=\sum_{\boldsymbol{i}}\left|\boldsymbol{i}-\boldsymbol{i}_{0}\right|^{2}C_{\boldsymbol{i}}\left(t\right).\label{eq:msd}
\end{equation}
For delocalized systems, the MSD grows as $t^{\beta_{\text{tr}}}$
with the exponent $\beta_{\text{tr}}$ characterizing the transport.
For example, $\beta_{\text{tr}}=1$ corresponds to diffusion, while
$\beta_{\text{tr}}=2$ corresponds to ballistic transport.

We argue that similar to the
surface roughness or the particle fluctuations, the MSD also follows
FV scaling, with a growth exponent $\beta_{\text{tr}}$. We define
delocalized systems, as system where an initial density excitation
spreads uniformly over the system, such that the MSD saturates to
$L^{2}$. Therefore, for delocalized systems, $\alpha_{\text{tr}}=2$,
and $z_{\text{tr}}=\alpha_{\text{tr}}/\beta_{\text{tr}}=2/\beta_{\text{tr}}$.
From the continuity equation, the change in the number of particles
in domain $A$ is related to the integrated current density in this
domain \citep{Derrida2009}. This means that the fluctuations in the
number of particles are related to the fluctuations of the integrated
current. The integrated current, is a local observable, which is connected
to a conserved quantity. As such, its fluctuations will saturate on
the time-scale it takes for a density perturbation to traverse the
subsystem. Given the above, we conjecture, that the saturation of
the particle number fluctuation occurs on the same time scale it takes
for a local density excitation to become uniform, $L^{z_{\text{tr}}}$.
This implies that $z=z_{\text{tr}}$, and since, as argued above,
$\alpha=1/2$ and $\alpha_{\text{tr}}=2$, it yields the dynamic scaling
relation, $\beta=\beta_{\text{tr}}/4$, which constitutes the main
result of this work. This relation is consistent with Ref.~\citep{fujimoto_dynamical_2021},
which found $\beta=1/4$ for a delocalized system with $\beta_{\text{tr}}=2$.
In what follows, we numerically verify this relation for one-dimensional
and three-dimensional non-interacting systems featuring \emph{distinct}
transport regimes, $\beta_{\text{tr}}$. We also provide evidence
that this relation holds in one-dimensional \emph{interacting} system.

\begin{figure}
\includegraphics[width=1\columnwidth]{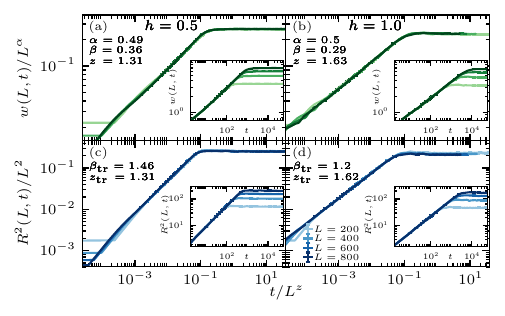}

\caption{Dynamical scaling of the surface roughness (top row) and the MSD (bottom
row) for the Fibonacci chain with $h=0.5,1.$ The inset shows the
data without the rescaling. System sizes used for the simulation are
$L=200-800$.}

\label{fig2}
\end{figure}

\section{Non-interacting systems}We consider a system of non-interacting
particles moving in an external potential with the Hamiltonian,
\begin{align}
\hat{H}= & -\sum_{\langle\boldsymbol{n},\boldsymbol{m}\rangle}J\left(\hat{a}_{\boldsymbol{n}}^{\dagger}\hat{a}_{\boldsymbol{m}}+\hat{a}_{\boldsymbol{m}}^{\dagger}\hat{a}_{\boldsymbol{n}}\right)+\sum_{\boldsymbol{n}}W_{\boldsymbol{n}}\hat{a}_{\boldsymbol{n}}^{\dagger}\hat{a}_{\boldsymbol{n}}.
\end{align}
Here $\hat{a}_{\boldsymbol{n}},\hat{a}_{\boldsymbol{n}}^{\dagger}$
annihilate and create a fermion at site $\boldsymbol{n}$ on a $d$-dimensional
lattice, $J$ is the hopping strength, which without loss of generality,
we set $J=1$, and $\langle.\rangle$ designates nearest neighbor
sites. The external potential is given by $W_{\boldsymbol{n}}$. We
focus on two specific systems: a one-dimensional Fibonacci chain of
length $L$ and a three-dimensional Anderson model on a cubical lattice
with side $L.$ For the Fibonacci chain, the onsite potential is given
by $W_{n}=h\left(2g\left(bn\right)-1\right)$, with $h$ being the
strength of the potential and $g(x)=\left[x+b\right]-\left[x\right]$
where $\left[x\right]$ denotes the integer part of $x$ and $b=\frac{1}{2}\left(\sqrt{5}-1\right)$
is the golden mean. The Fibonacci model exhibits multi-fractal single-particle
states for \emph{any} value of the potential strength~\citep{ostlund_one-dimensional_1983,kalugin_electron_1986,luck_phonon_1986,kohmoto_critical_1987,niu_spectral_1990}.
Transport is known to crossover from ballistic transport $\left(\beta_{\text{tr}}=2\right)$
at $h=0$ to sub-diffusive transport $\left(\beta_{\text{tr}}<1\right)$
for large $h$ \citep{Hiramoto1988dynamics,piechon1996anomalous,varma_fractality_2017,varma_diffusive_2019,chiaracane_quantum_2021}.
For the Anderson model in three-dimension, the onsite potential is
uniformly drawn from the interval $W_{\boldsymbol{n}}\in\left[-h/2,h/2\right]$,
where $h$ corresponds to the strength of the disorder. The delocalized
phase of the Anderson model is diffusive ($\beta_{\text{tr}}=1$),
and anomalous transport ($\beta_{\text{tr}}=2/3$) and multi-fractal
eigenstates exist only at the critical point, $h=16.5J$ \citep{tomi_ohtsuki1997,vasquez_multifractal_2008,rodriguez_multifractal_2008,evers_anderson_2008}.
We average the quantities of interest over $50$ independent disorder
realizations for the Anderson modeland over a \emph{$50$ }samples
of the Fibonacci sequence for the Fibonacci chain~\footnote{By generating a Fibonacci sequence with $N\gg L$ terms, we can cut
out $L+1$ independent samples of of length $L$. Removing sequences
which are related by reflection leads to $\frac{L}{2}$ or $\left[\frac{L-1}{2}\right]$
independent configurations for even {[}odd{]} number of lattice sites.
~\citep{varma_diffusive_2019,chiaracane_quantum_2021}.}.

To maximize the growth of the surface roughness, we initiate the system
from states with a definite number of particles in the domain $A$,
which we chose to be half of the system (see Fig.~\ref{fig1}). For
the Fibonacci chain, we evolve the system starting from a charge-density-wave
state, $\left|\psi_{\text{1D}}\right\rangle =\prod_{i=1}^{L/2}\hat{a}_{2i}^{\dagger}\left|0\right\rangle $
and for the three-dimensional Anderson model, we use the checkerboard
pattern, $\left|\psi_{\text{3D}}\right\rangle =\prod_{\text{mod}(i+j+k,2)=0}\hat{a}_{i,j,k}^{\dagger}\left|0\right\rangle .$
Here $\left|0\right\rangle $ denotes the vacuum and $\left(i,j,k\right)$
correspond to a point on a cubical lattice in 3D. The energy of these
states is close to the middle of the many-body spectrum, therefore
for delocalized dynamics we expect the surface roughness to approach
the infinite temperature limit. The surface roughness is calculated
after numerically evolving the single-particle density matrix. This
matrix is dense and and has $L^{2d}$ elements, therefore, for the
3D Anderson problem we are limited to linear dimension of about $L=24$.

\begin{figure}
\includegraphics[width=1\columnwidth]{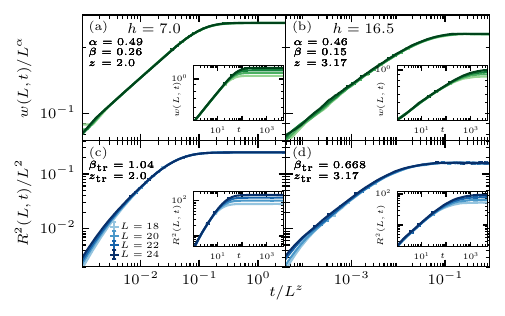}\caption{Same as Fig.~\ref{fig2} but for the 3D Anderson model at the delocalized
$\left(h=7\right)$ and critical $\left(h=16.5\right)$ phases. The
linear dimension of the system sizes we simulated are $L=18-24$.}

\label{fig3}
\end{figure}

\begin{figure}
\includegraphics[width=1\columnwidth]{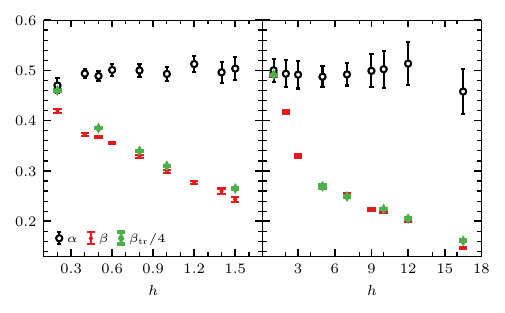}

\caption{The open black circles correspond to the $\alpha$ exponent, and the
full red circles to the $\beta$ exponent of the surface roughness
as a function of the potential strength $h$. The left panel is the
Fibonacci chain and the right column is the three-dimensional Anderson
model. The green diamonds correspond to $\beta_{\text{tr}}/4$, where
$\beta_{\text{tr}}$ is the transport exponent as computed from the
MSD.}

\label{fig4}
\end{figure}

The insets of Figs.~\ref{fig2}(a,b) show the dynamics of the surface
roughness for the Fibonacci model for a set of parameters $h=0.5,1.0$,
and a range of system sizes $L=200-800$, while the insets of Figs.
\ref{fig3}(a-b) show a similar plot for the three-dimensional Anderson
model for the delocalized $\left(h=7\right)$ and critical $\left(h=16.5\right)$
phases, and for a linear dimension of $L=18-24$. The surface roughness
grows as a power-law in time and saturates to a system size dependent
value. The data for different system sizes can be collapsed into a
single curve by rescaling $w\left(L,t\right)$ and $t$ by $L^{\alpha}$
and $L^{z}$, respectively {[}see main panels of Figs.~\ref{fig2}(a,b)
and \ref{fig3}(a-b){]}, which suggests the existence of the FV scaling.
We perform the scaling collapse by minimizing the deviation $\chi\left(\alpha,z\right)=\sum_{L,t}\left|w\left(L_{0},t\right)-\left(L/L_{0}\right){}^{-\alpha}w\left(L,\left(L/L_{0}\right)^{z}t\right)\right|/w^{2}\left(L_{0},t\right)$,
with $L_{0}$ being a reference system size, which allows us to extract
the exponents $\alpha$ and $\beta$ \citep{pineiro_orioli_universal_2015,fujimoto_family-vicsek_2020}
(See appendix~\ref{sec:Multifractality-and-dynamical} for the $z$-exponents). Fig.~\ref{fig4}
shows the calculated exponents for both systems as a function of the
external potential. For all values of the external potential, $\alpha\approx1/2$
and $\beta$ is monotonously decreasing from a ballistic value $\left(\beta=1/2\right)$.
Interestingly, contrary to the predictions of the classical EW equation
in three dimensions predicts $\left(\alpha,\beta<0\right)$~\citep{edwards_surface_1982,Nattermann1992kinetic},
we find that the quantum surface roughness exhibits exponents similar
to that of the one-dimensional case. The localized phase do not show
FV scaling since the surface roughness saturates to a value which
does not depend on the system size (data not shown). We now proceed
showing that the MSD also follows the FV scaling and that the dynamical
scaling relation, $\beta=\beta_{\text{tr}}/4$ holds.

We numerically obtain the MSD by propagating the correlation function
(\ref{eq:corr_function}), which for non-interacting systems, amounts
to evolving a a vector of length $L^{d}$. As such, much larger system
sizes are accessible compared to the surface roughness. Since the
initial charge-density wave initial conditions we use for the surface
roughness correspond to an energy density in the middle of the many-body
spectrum, we compute the correlation function at infinite temperate.
In Figs.~\ref{fig2}(c,d) and \ref{fig3}(c,d) we show that similarly
to the surface roughness, the MSD follows FV scaling with the exponents
$\alpha_{\text{tr}}=2$ and $\beta_{\text{tr}}$. Fig.~\ref{fig4}
shows a remarkable agreement between $\beta_{\text{tr}}/4$ and $\beta$
for both the Fibonacci chain and the 3D Anderson models for all studied
strengths of the potential, verifying the dynamic scaling relation.
For both models the $\beta_{\text{tr}}/4$ is monotonically decreasing
from a ballistic value for very weak disorder $\left(\beta_{\text{tr}}/4=1/2\right)$
to sub-diffusive values $\left(\beta_{\text{tr}}/4<1/4\right)$ for
higher potential strengths. We note that, for the 3D Anderson model,
diffusion is expected $\left(\beta_{\text{tr}}/4=1/4\right)$ for
$h<h_{c}=16.5$. However, its numerical observation requires $l\ll L$,
where $l$ is the mean-free path. Since $l\sim h^{-2}$, large systems
sizes are required to observe the asymptotic transport for weak external
potential (See appendix ~\ref{sec:Finite-size-analysis} for finite size analysis).
Interestingly, even outside the asymptotic transport regime the relationship
between $\beta_{\text{tr}}$ and $\beta$ is satisfied.
\begin{figure}
\includegraphics[width=1\columnwidth]{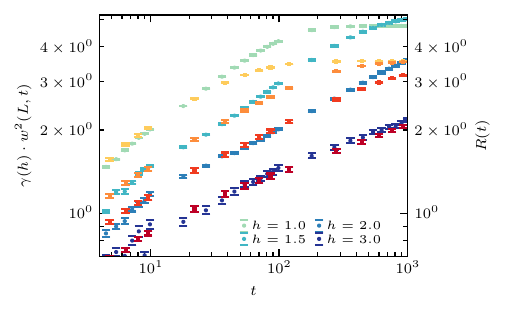}

\caption{The square of surface roughness, $w^{2}\left(L,t\right)$ (blue points)
and root mean-squared displacement, $R\left(t\right)$ (red squares)
as a function of time and a number of potential strengths for the
interacting Fibonacci chain with parameters $L=24,\ V=1.0$. More
intense colors correspond to larger potential strength. The surface
roughness is multiplied by a disordered dependent factor $\gamma(h)$
to obtain a visual match with $R\left(t\right)$ at early times.}

\label{fig5}
\end{figure}

\section{Interacting system} To verify if the dynamic scaling relationship
holds in the presence of interactions we consider an interacting Fibonacci
chain with the nearest-neighbor interaction
\begin{equation}
\hat{H}_{\text{int}}=V\sum_{i}{\displaystyle \left(\hat{n}_{i}-1/2\right)\left(\hat{n}_{i+1}-1/2\right)},
\end{equation}
where $V$ is the interaction strength. We focus on potential strengths
for which the system is ergodic \citep{mace_many-body_2019,varma_diffusive_2019,chiaracane_quantum_2021}.
For the interacting case, the numerical complexity is exponential
in the system size, such that we are limited to $L=24$ sites. Due
to the limited power-law growth regime it is hard to reliably extract
$\beta$ and $\beta_{\text{tr }}$ for such system sizes (see appendix ~\ref{sec:Finite-size-analysis}
for finite size analysis). Instead, in Fig.\ref{fig5}, we compare
the root mean-squared displacement $R(t)$ to the square of the surface
roughness $w^{2}(L,t)$. If the dynamic relation, $\beta=\beta_{\text{tr}}/4$,
holds, these quantities are supposed to be proportional up to a constant
dependent on the potential strength. In Fig.~\ref{fig5}, we see
that this is indeed the case: both quantities have the same growth
exponent, though, the surface roughness takes more time to reach saturation.
While we focus on half-filling here, we show in appendix~\ref{sec:dynamics-away-from-half-filled},
that such a relation also holds away from the half-fillings as well.

\section{Discussion}. In this paper we conjecture that mean-square
displacement (MSD) follows the Family-Vicsek (FV) scaling, and saturates
on the same time scale as the surface roughness, or analogously, particle
number fluctuations. Using this conjecture, we obtain a dynamic relationship
between the transport exponent, and the growth exponent of the surface
roughness, $\beta=\beta_{\text{tr}}/4$, which applies to any particle
density.We numerically confirm this conjecture by studying two prototypical
non-interacting quantum systems in one and three dimensions, and one
interacting system in one dimension, where transport can be controlled
by the strength of the external potential. While we numerically confirm
the dynamical relationship in fermionic systems, since our argument
does not depend on the statistics of the particles, we expect it to
also hold for bosonic systems.

It is important to note, that for \emph{classical diffusive and non-interacting}
systems, a connection between particle fluctuations and transport,
can be derived using fluctuating hydrodynamics, which puts diffusive
systems in the Edwards-Wilkinson (EW, $\beta=1/4$) universality class
~\citep{Kipnis1982,krapivsky2012fluctuations,eldad2022inverse}.
However, for simple \emph{interacting} cases, such as the symmetric
exclusion process, this relation does \emph{not} hold~\citep{alexander1978diffusion,Derrida2009,Derrida2009ssep,barkai2010diffusion}.
Moreover, there is no known relationship between these quantities
for \emph{classical} systems with \emph{anomalous} transport. On the
contrary, our results suggest that for \emph{quantum} systems, the
dynamic scaling relationship holds also for anomalous transport and
interacting systems. We argue that the difference between classical
and quantum cases follows due to indistinguishably between the particles
in quantum dynamics, which allows collective relaxation of the density
even in one-dimensional systems, where classical dynamics exhibit
single-file motion. In higher dimensions, we expect that the dynamical
scaling relation will hold also for classical interacting systems.

The established dynamic scaling relationship explains the mechanism
behind the quantum surface roughness growth, and provides a convenient
way to assess transport in cold-atoms experiments, where particle
number fluctuations can be directly measured. Indeed, while our work
was in preparation, this method was used in a recent cold-atoms experiment~\citep{wienand2023emergence}.

Our study leaves a number of questions open. For non-interacting systems
the transport exponent is related to the multi-fractal properties
of the single-particles states \citep{ketzmerick_slow_1992,ketzmerick_what_1997,tomi_ohtsuki1997}
(see also appendix ~\ref{sec:Multifractality-and-dynamical}). Is there a similar relationship
for the interacting system? Can one derive a formal connection between
surface roughness growth and transport? Is the dynamic relation sensitive
to the energy density of the initial state? And how the universality
class changes when the system is coupled with a dissipative environment?

During the final stages of preparation of this manuscript a related
and complimentary study has appeared \citep{aditya2023family}.

\begin{acknowledgments}
We would like to thank Naftali Smith for fruitful discussions. This
research was supported by a grant from the United States-Israel Binational
Foundation (BSF, Grant No. $2019644$), Jerusalem, Israel, and the
United States National Science Foundation (NSF, Grant No. DMR$-1936006$),
and by the Israel Science Foundation (grants No. 527/19, 218/19 and
1304/23). D.S.B acknowledges funding from the Kreitman fellowship.
\end{acknowledgments}

\bibliography{ref}

\onecolumngrid
\appendix

\section{Particle number fluctuations and surface roughness}\label{particle-number-roughness}

Particle number in domain $A$ are defined in  Eq.2
in the main text,
\begin{align}
	\hat{N}_{A}= & \sum_{\boldsymbol{i}\in A}\hat{n}_{\boldsymbol{i}}\left(t\right),
\end{align}
which can be contrasted to the height operator defined in Refs.~\citep{fujimoto_family-vicsek_2020,fujimoto_dynamical_2021},
\begin{align*}
	\hat{h}_{A}= & \sum_{\boldsymbol{i}\in A}\left(\hat{n}_{\boldsymbol{i}}\left(t\right)-\nu\right)=\hat{N}_{A}-\nu V_{A},
\end{align*}
where $\nu=N/V$ is the particle density, and $V_{A}$ is the number
of sites in domain $A$. Since $\nu V_{A}$ is a constant, the fluctuations
in $\hat{N}_{A}$ coincide with the surface roughness -- the fluctuations
of the height operator.

At infinite temperature the average number of particles in domain
$A$ is,
\begin{equation}
	\left\langle \hat{N}_{A}\right\rangle _{\infty}=\sum_{\boldsymbol{i}\in A}\left\langle \hat{n}_{\boldsymbol{i}}\right\rangle _{\infty}=\nu V_{A}.
\end{equation}
Similarly, the square of the number of particles is,
\begin{align}
	\left\langle \hat{N}_{A}^{2}\right\rangle _{\infty} & =\sum_{\boldsymbol{i}\in A}\sum_{\boldsymbol{j}\in A}\left\langle \hat{n}_{\boldsymbol{i}}\hat{n}_{\boldsymbol{j}}\right\rangle _{\infty}=\sum_{\boldsymbol{i}\in A}\left\langle \hat{n}_{\boldsymbol{i}}^{2}\right\rangle _{\infty}+\sum_{\boldsymbol{i}\neq\boldsymbol{j}\in A}\left\langle \hat{n}_{\boldsymbol{i}}\hat{n}_{\boldsymbol{j}}\right\rangle _{\infty}\nonumber \\
	& =\sum_{\boldsymbol{i}\in A}\left\langle \hat{n}_{\boldsymbol{i}}^{2}\right\rangle _{\infty}+\sum_{\boldsymbol{i}\neq\boldsymbol{j}\in A}\left\langle \hat{n}_{\boldsymbol{i}}\right\rangle \left\langle \hat{n}_{\boldsymbol{j}}\right\rangle _{\infty}=\alpha_{\nu}V_{A}+\nu^{2}V_{A}\left(V_{A}-1\right),
\end{align}
where $\alpha_{\nu}\equiv\left\langle \hat{n}_{\boldsymbol{i}}^{2}\right\rangle $
is a constant which is $\nu$ for fermions and $\nu\left(1+\nu\right)$
for bosons. To simplify the calculation of $\left\langle \hat{n}_{i}\hat{n}_{j}\right\rangle $,
we have used the grand-canonical ensemble, where $\left\langle \hat{n}_{\boldsymbol{i}}\hat{n}_{\boldsymbol{j}}\right\rangle _{\infty}=\left\langle \hat{n}_{\boldsymbol{i}}\right\rangle \left\langle \hat{n}_{\boldsymbol{j}}\right\rangle _{\infty}$
for $i\neq j$. Therefore, the fluctuations at infinite temperature
are given by,
\begin{equation}
	\Delta N_{A}=\sqrt{\left\langle \hat{N}_{A}^{2}\right\rangle _{\infty}-\left\langle \hat{N}_{A}\right\rangle _{\infty}^{2}}=\sqrt{\left(\alpha_{\nu}-\nu^{2}\right)V_{A}}=\begin{cases}
		\sqrt{\nu\left(1-\nu\right)V_{A}} & \text{fermions}\\
		\sqrt{\nu V_{A}} & \text{bosons}
	\end{cases}.
\end{equation}
For a domain half the size of the system $V_{A}=L^{d}/2$ and therefore,
\begin{equation}
	\left\langle \hat{N}_{A}\right\rangle _{\infty}\sim L^{d}\qquad\Delta N_{A}\sim L^{d/2}\qquad w_{\infty}\left(L,t\right)\sim L^{d/2-\left(d-1\right)/2}=L^{1/2}.
\end{equation}
A similar calculation can be performed also for a fixed number of
particles with canonical ensemble. To leading order in $1/V$, it
gives,
\begin{equation}
	\left\langle \hat{n}_{i}\hat{n}_{j}\right\rangle =\frac{1}{4}-\frac{1}{4V},
\end{equation}
which yields a sub-leading, and therefore insignificant, correction
to $w_{\infty}\left(L,t\right)$.

\section{Dynamics of entanglement entropy}\label{entropy}

For a sub-system described by a reduced density matrix $\rho_{s}(t)$,
the dynamics of the entanglement entropy can be calculated as: $S(L,t)=-\text{Tr}[\rho_{s}(t)\text{ln}\rho_{s}(t)]$,
which for the free fermionic case, is related to the eigenvalues $c_{\alpha}$
of the two-point correlation function $\langle\hat{a}_{i}^{\dagger}(t)\hat{a}_{j}(t)\rangle$
restricted to the subsystem \citep{peschel_calculation_2003,peschel_reduced_2009},
\begin{align}
	S(L,t) & =-\sum_{i}\left[c_{\alpha}\log c_{\alpha}+\left(1-c_{\alpha}\right)\log\left(1-c_{\alpha}\right)\right].
\end{align}

For non-interacting systems, the entanglement entropy is related to
the surface roughness or the particle number fluctuations in the subsystem
as $S(L,t)\propto w^{2}(L,t)$ \citep{klich_quantum_2009,fujimoto_dynamical_2021}.
Thus, the dynamics of the entanglement entropy is also expected to
follow the FV scaling with the modified exponents $(2\alpha,2\beta,z)$.

\begin{figure}
	\includegraphics[width=12cm]{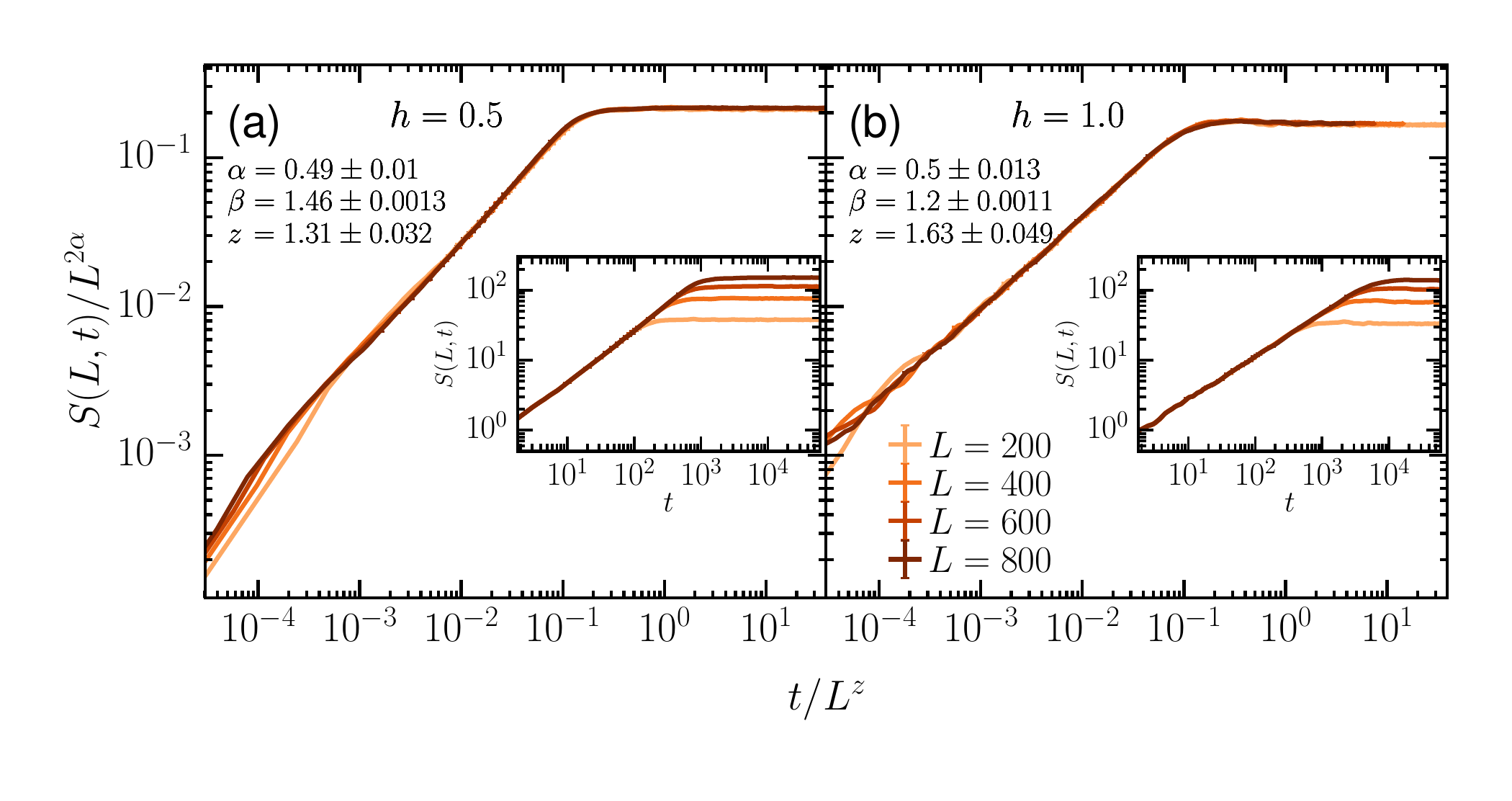}
	
	\caption{Dynamical scaling of the entanglement entropy for the Fibonacci chain
		with $h=0.5,1.$ The inset shows the data without the rescaling. System
		sizes used for the simulation are $L=200-800$.}
	
	\label{fig6}
\end{figure}

\begin{figure}
	\includegraphics[width=12cm]{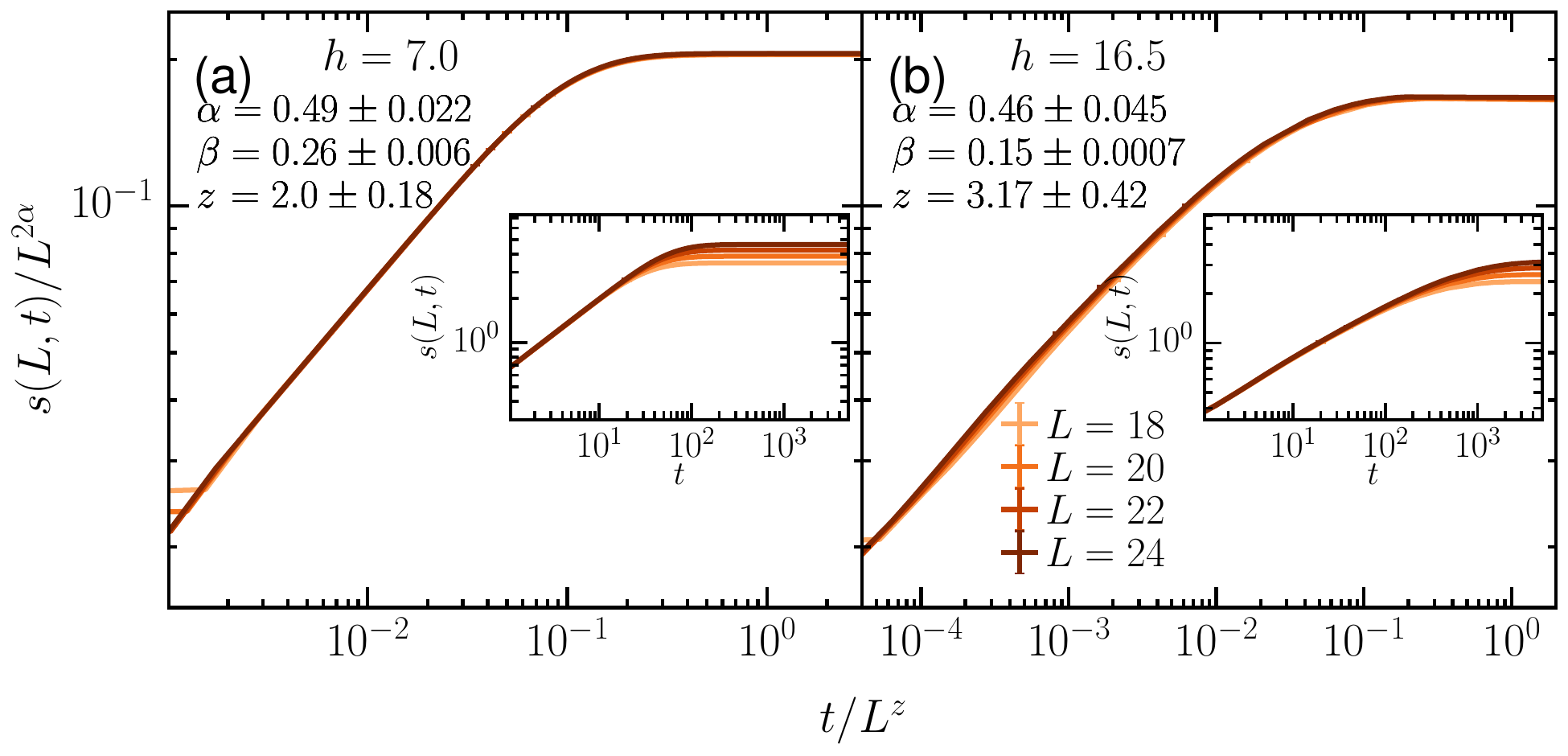}
	
	\caption{Same as Fig.~\ref{fig6} but for the 3D Anderson model at the delocalized
		$\left(h=7\right)$ and critical $\left(h=16.5\right)$ phases. The
		linear dimension of the system sizes we simulated are $L=18-24$.}
	
	\label{fig7}
\end{figure}

\begin{figure}
	\includegraphics[width=12cm]{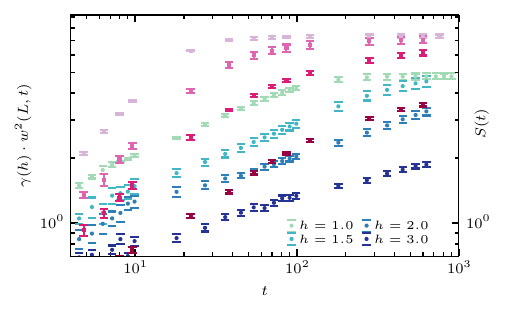}
	
	\caption{Dynamics of entanglement entropy (red squares) for the interacting
		Fibonacci chain for $L=24$. The square of surface roughness (blue
		points), $w^{2}\left(L,t\right)$ is also plotted for a comparison.}
	
	\label{fig_ent}
\end{figure}

We plot the dynamics of the half-chain entanglement entropy in Fig.~\ref{fig6}
for the non-interacting Fibonacci chain for potential strengths $h=0.5,\:1.0$,
and for system sizes $L=200-800$. Similar to the surface roughness,
the entanglement entropy grows in a power-law fashion followed by
a system size dependent saturation (insets of Fig. \ref{fig6}(d-f)).
Performing a rescaling of $S(L,t)$ and $t$ by $L^{2\alpha}$ and
$L^{z}$ respectively, we see a collapse of the data for different
$L$ to a single curve (Fig. \ref{fig6}) which suggest the presence
of FV scaling in the entanglement entropy.

Similarly, for the three dimensional Anderson model, the entanglement
entropy normalized by the area: $s\equiv$$S/L^{2},$ is plotted in
(Fig.~\ref{fig7}) for disorder strengths $h=7,\ 16.5$ and linear
dimensions $L=18-24$. We consider the subsystem to be one half of
the cube and the normalization to remove the dependence of the system
size on the initial dynamics. Similarly to the one-dimensional Fibonacci
chain, we see the presence of FV scaling in the entanglement entropy
for the three dimensional Anderson model with the exponents $(2\alpha,2\beta,z)$.
For the interacting case, the relationship between the particle number
number fluctuation and entanglement entropy does not hold anymore
and the exponents $(2\alpha,2\beta,z)$ are no longer expected. However,
particle number fluctuation present a lower bound on the entanglement
entropy, such that $\beta_{\text{ent}}>\beta$ \citep{klich_quantum_2009,kiefer-emmanouilidis_bounds_2020}.
We confirm this by studying the dynamics of the entanglement entropy
for the interacting Fibonacci chain. We plot the dynamics of the entanglement
entropy for the system size $L=24$, and a range of potential strengths
$h=1.0-3.0$ in Fig.~\ref{fig_ent}. Similarly to the surface roughness
and the RMSD, we see that the entanglement entropy grows as a power-law
in time followed by a saturation. A comparison with the square of
the surface roughness/particle number fluctuations $w^{2}\left(L,t\right)$
is also provided in Fig.~\ref{fig_ent}. In contrast to the non-interacting
case, the entanglement entropy grows faster than $w^{2}\left(L,t\right)$
and the relationship $\beta_{\text{ent}}=2\beta$ does not hold anymore.

\section{Finite size analysis\protect\label{sec:Finite-size-analysis}}

\begin{figure}
	\includegraphics[width=8cm]{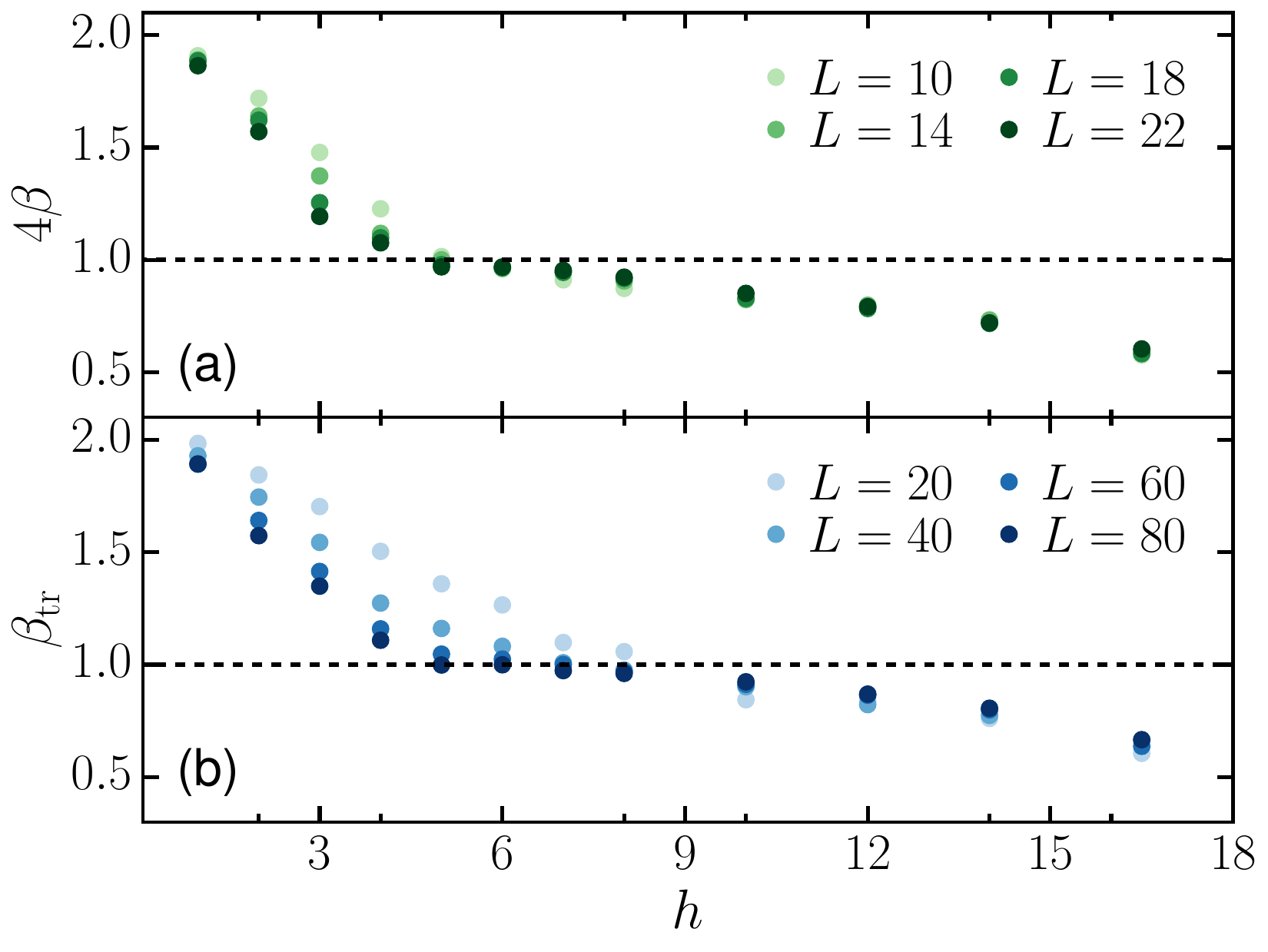}
	
	\caption{Finite size analysis for the three-dimensional Anderson model. The
		top panel is the growth exponent for $L=14-22$. The bottom panel
		is for transport growth for $L=20-80$. The flow towards the EW class
		is clearly visible in the delocalized side. The dashed line corresponds
		to the diffusive transport with $\beta_{\text{tr}}=1.$}
	
	\label{fig8}
\end{figure}

In Fig. \ref{fig8}, we plot the surface roughness exponent $\beta$
and the transport exponent $\beta_{\text{tr}}$ for the $3$D Anderson
model, as a function of the potential strength $h$ system sizes $L=14-22.$
The surface roughness exponent is multiplied by a factor of $4$ to
match with the transport exponent. It can be seen that for all the
disorder strengths, the relationship between the exponents $\beta_{\text{tr}}=4\beta$
holds. Furthermore, for disordered smaller than the critical disorder
strength $(h<h_{c}=16.5)$, we see that both $4\beta$ and $\beta_{\text{tr}}$
approach to $1$ (marked by a dashed line) as the system size increases
and signifies diffusive transport. These growth exponents suggest
a EW universality class for quantum diffusive systems.

\begin{figure}
	\includegraphics[width=8cm]{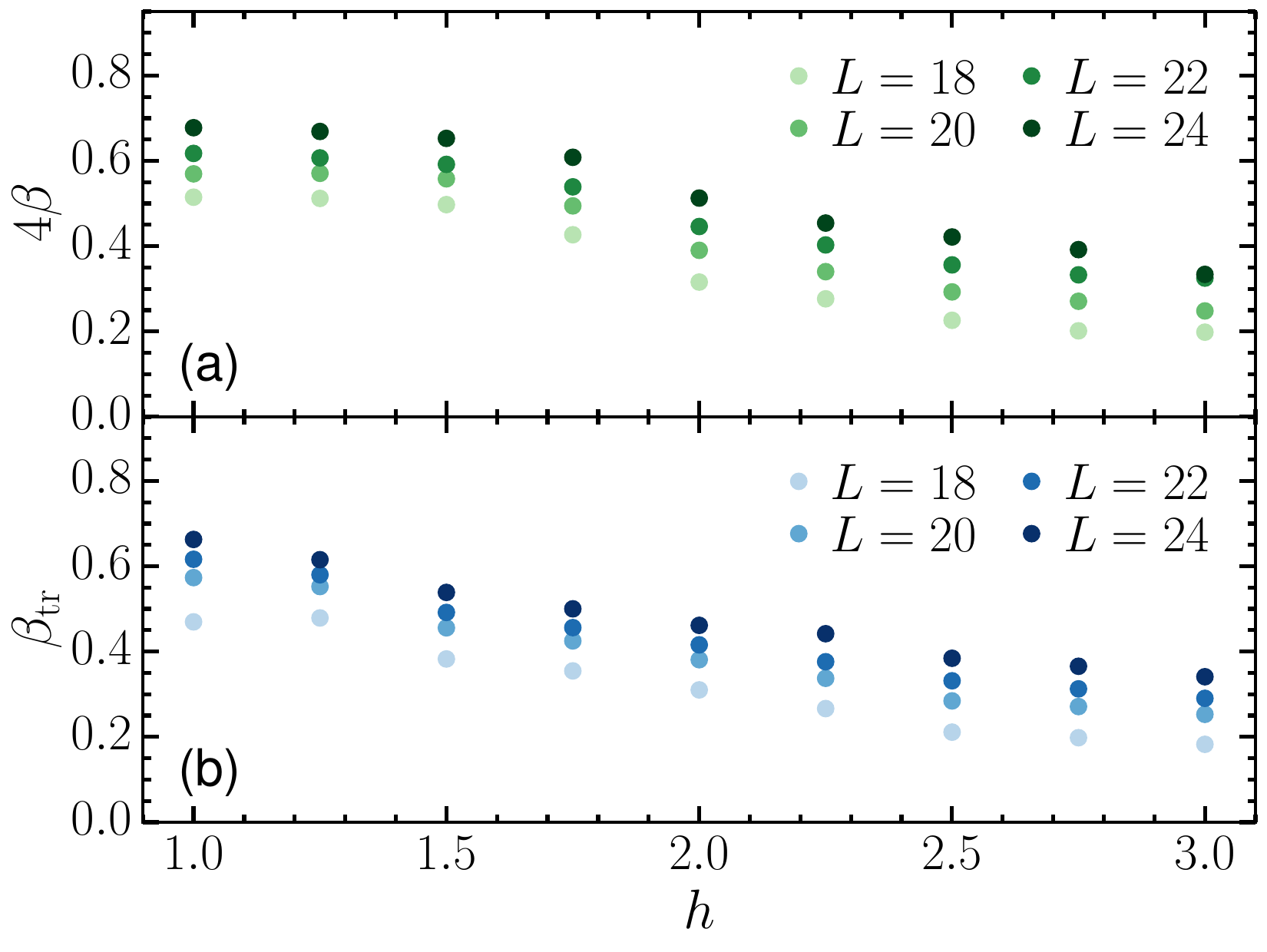}
	
	\caption{Similar to Fig.~\ref{fig8}, but for the interacting Fibonacci chain
		with $V=1.0$. The system sizes are $L=18-24$.}
	
	\label{fig9}
\end{figure}

In Fig.~\ref{fig9} we present a similar analysis for the interacting
Fibonacci chain for interaction strength $V=1$ and system sizes $L=18-24$.
Similar to the non-interacting case, we see that the relationship
between the transport exponent $\beta_{\text{tr}}$and the surface
roughness/particle fluctuations exponent $\beta$ holds $(\beta_{\text{tr}}=4\beta)$
. 

\section{Multifractality and dynamical exponents\protect\label{sec:Multifractality-and-dynamical}}

Eigenstates of quantum systems often exhibit self-similar (multi-)fractal
behavior in the vicinity of a quantum phase transition (QPT) \citep{huckestein1995scaling,evers_multifractality_2001,subramaniam2006surface,evers_anderson_2008,evers_multifractality_2008,tikhhonov2018many,mace2019multifractal,solorzano2021multifractal}.
Such states are spatially extended, but sparse, and are characterized
by (multi-)fractal dimensions. A celebrated example of multifractal
eigenstates occurs at the critical point of the $3$D Anderson model~\citep{anderson_absence_1958,abrahams_scaling_1979,evers_anderson_2008,lee_disordered_1985,kramer_localization_1993,grussbach_determination_1995,edwards_numerical_1972}.
Other notable examples are ground state of quantum Hall systems~\citep{evers_multifractality_2001,evers_multifractality_2008},
quasiperiodic systems~\citep{aubry1980analyticity,roy_multifractality_2018,deng_one-dimensional_2019,sarkar_mobility_2021},
long-range systems \citep{parshin_multifractal_1998,cuevas_anomalously_2001,mirlin_transition_1996,varga_critical_2000},
and random regular graphs~\citep{altshuler_nonergodic_2016}. Using
the previously established connection between transport in non-interacting
systems and (multi)fractality of single-particle eigenstates \citep{ketzmerick_slow_1992,ketzmerick_what_1997,tomi_ohtsuki1997},
we argue that the surface growth exponent is also related to (multi)fractal
properties.

\begin{figure}
	\includegraphics[width=8cm]{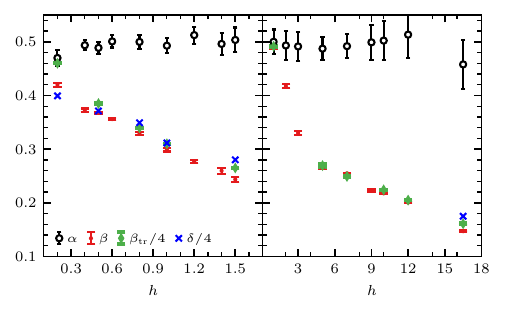}\includegraphics[width=8cm]{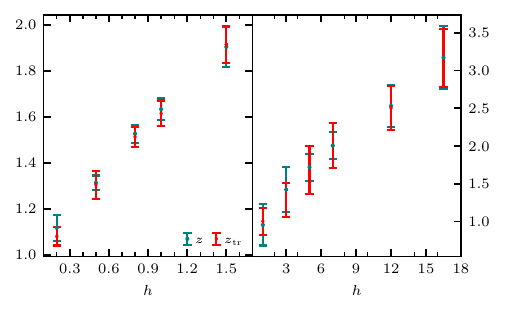}
	
	\caption{\emph{Left}: Dynamic exponents of the surface roughness $\alpha$
		(open black circles), $\beta$ (red circles) and transport exponent
		$\beta_{\text{tr}}$ (green diamonds) as a function of the potential
		strength $h$ for the Fibonacci chain (left) and the three-dimensional
		Anderson model (right). The black crosses correspond to $\delta/4$,
		where $\delta=D_{2}^{\mu}/D_{2}^{\psi}$. \emph{Right}: Dynamical
		exponents $z$ (circles) and $z_{\text{tr}}$ (diamonds) as a function
		of $h$ for Fibonacci and $3$D Anderson model.}
	
	\label{fig10}
\end{figure}

For non-interacting systems, the dynamical exponent of the MSD growth
$\beta_{\text{tr}}$ is related to the fractal dimensions, $\delta=D_{2}^{\mu}/D_{2}^{\psi}$
\citep{ketzmerick_slow_1992,ketzmerick_what_1997,tomi_ohtsuki1997}.
Here $D_{2}^{\mu}$ is the correlation dimension of the local density
of states and $D_{2}^{\psi}$ is the correlation dimension of the
single-particle eigenstates. Since we have shown that the transport
exponent is related to the surface roughness growth exponent as $\beta=\beta_{\text{tr}}/4$,
using the relationship $\beta_{\text{tr}}=\delta=D_{2}^{\mu}/D_{2}^{\psi}$
we obtain $\beta\simeq\delta/4$. We verify this relation for both
the Fibonacci and Anderson models by plotting $\beta$ and $\delta/4$
for various values of $h$ (see Fig.~\ref{fig10}) using the values
of $D_{2}^{\mu}$ and $D_{2}^{\psi}$ from Refs.~\citep{ketzmerick_slow_1992,ketzmerick_what_1997,tomi_ohtsuki1997}.

\section{Dynamics away from half-filling\protect\label{sec:dynamics-away-from-half-filled}}

\begin{figure}[h]
	\includegraphics[width=8cm]{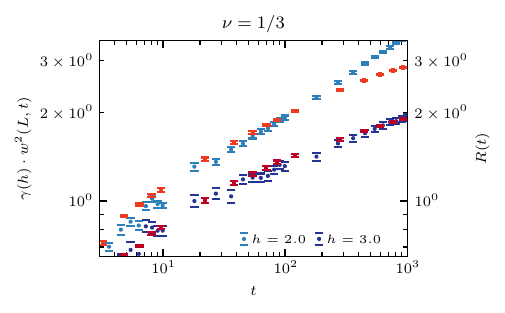}\includegraphics[width=8cm]{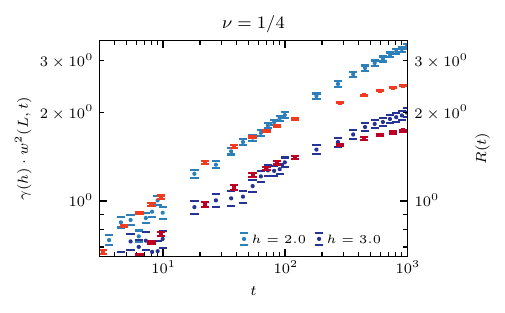}
	
	\caption{The square of surface roughness, $w^{2}\left(L,t\right)$ (blue points)
		and root mean-squared displacement, $R\left(t\right)$ (red squares)
		as a function of time and for different potential strengths for the
		interacting Fibonacci chain with parameters $L=24,\ V=1.0$ and filling
		factor $\nu=1/3$ (left), and $\nu=1/4$ (right). More intense colors
		correspond to larger potential strength. The surface roughness is
		multiplied by a disordered dependent factor $\gamma(h)$ to obtain
		a visual match with $R\left(t\right)$ at early times.}
	
	\label{fig12}
\end{figure}

In this section, we study the dynamics of the particle number fluctuations
and the transport away from the half-filling. We specifically focus
on the interacting case and perform the same analysis presented in
Fig.\ref{fig5} but with filling factor $\nu=1/3$ and $\nu=1/4$
respectively for a system size $L=24$ (Fig.\ref{fig12}). Similar to the half-filling
case, we see that the relationship between the dynamical exponents
also holds away from the half-filling.

\end{document}